\documentclass[aip,jcp,reprint,eqsecnum,superscriptaddress,twocolumn]{revtex4-1}

\usepackage{natbib}
\usepackage{amssymb,amsmath}

\usepackage{epsfig}
\usepackage{bm}
\usepackage{color}
\usepackage{graphicx}
\usepackage{hyperref}
\usepackage{enumerate}

\newcommand\beq{\begin{equation}}
\newcommand\eeq{\end{equation}}
\newcommand\beqa{\begin{eqnarray}}
\newcommand\eeqa{\end{eqnarray}}
\newcommand{\nn}{\nonumber\\}
\def\bal#1\eal{\begin{align}#1\end{align}}

\newcommand{\ma}{m_i}

\newcommand{\mb}{m_j}

\newcommand{\cc}{\mathbf{v}}

\newcommand{\kk}{\widehat{\bm{\sigma}}}

\newcommand{\ww}{\bm{\omega}}

\newcommand{\Ia}{I_i}

\newcommand{\da}{\sigma_i}

\newcommand{\db}{\sigma_j}

\newcommand{\dab}{\sigma_{ij}}

\newcommand{\x}{\times}

\newcommand{\een}{\alpha_{ij}}

\newcommand{\eet}{\beta_{ij}}

\newcommand{\en}{\overline{\alpha}_{ij}}

\newcommand{\et}{\overline{\beta}_{ij}}

\newcommand{\mab}{m_{ij}}

\newcommand{\qab}{\kappa_{ij}}

\newcommand{\qa}{\kappa_{i}}

\newcommand{\qb}{\kappa_{j}}

\newcommand{\q}{\kappa}

\newcommand{\fa}{f_{i}}

\newcommand{\fb}{f_{j}}

\newcommand{\far}{f_{i}^{\text{rot}}}

\newcommand{\Tat}{T_{i}^{\text{tr}}}

\newcommand{\Tbt}{T_{j}^{\text{tr}}}

\newcommand{\Tt}{T^{\text{tr}}}

\newcommand{\Tar}{T_{i}^{\text{rot}}}

\newcommand{\Tbr}{T_{j}^{\text{rot}}}

\newcommand{\Tr}{T^{\text{rot}}}

\newcommand{\Qab}{J_{ij}}

\newcommand{\na}{n_i}

\newcommand{\nb}{n_j}

\newcommand{\zabt}{\xi_{ij}^{\text{tr}}}

\newcommand{\zt}{\xi^{\text{tr}}}

\newcommand{\zabr}{\xi_{ij}^{\text{rot}}}

\newcommand{\zr}{\xi^{\text{rot}}}

\newcommand{\al}{i}

\newcommand{\be}{j}

\newcommand{\tr}{{\text{tr}}}

\newcommand{\rot}{{\text{rot}}}

\newcommand{\wwwab}{\mathbf{w}_{ij}}

\begin{document}

\title{Intruders in disguise: Mimicry effect in granular gases}

\author{Antonio Lasanta}
\email{alasanta@ing.uc3m.es}
\affiliation{Gregorio Mill\'an Institute of Fluid Dynamics,
Nanoscience and Industrial Mathematics,
Department of Materials Science and Engineering and Chemical Engineering,
Universidad Carlos III de Madrid, 28911 Legan\'es, Spain}

\author{Francisco Vega Reyes}
\email{fvega@unex.es}

\author{Vicente Garz\'o}
\email{vicenteg@unex.es}
\author{Andr\'es Santos}
\email{andres@unex.es}

\affiliation{Departamento de F\'{\i}sica and Instituto de
  Computaci\'on Cient\'{\i}fica Avanzada (ICCAEx), Universidad de
  Extremadura, 06006 Badajoz, Spain}

\begin{abstract}
In general, the total kinetic energy in a multicomponent granular gas of inelastic and rough hard spheres  is unequally partitioned among the different degrees of freedom. On the other hand, partial energy equipartition can be reached, in principle, under appropriate combinations of the mechanical parameters of the system. Assuming common values of the coefficients of restitution,  we use kinetic-theory tools to determine the conditions under which the components of a granular mixture in the homogeneous cooling state have the same translational and rotational temperatures as those of a one-component granular gas (``mimicry'' effect). Given the values of the concentrations and the size ratios, the mimicry effect requires the mass ratios to take specific values, the smaller spheres having a larger particle mass density than the bigger spheres. The theoretical predictions for the case of an impurity immersed in a host granular gas are compared against both DSMC and molecular dynamics simulations with a good agreement.

\end{abstract}

\date{\today}
\maketitle
\section{Introduction}

\label{sec1}

Graeme Austin Bird  was a gigantic figure in the field of rarefied gas dynamics. He developed an intuitive and original stochastic algorithm---the direct simulation Monte Carlo (DSMC) method---that obtains exact numerical solutions of the Boltzmann equation.\cite{B94,B13} The DSMC method boosted a fruitful new area of research with many important applications in science and engineering alike.\cite{GTR01,GTPRK14,GC17,RPMY18,CKAM19,AV19}
This technique was later imported
into the field of granular gas dynamics (gases of macroscopic particles that
undergo inelastic collisions), where the total kinetic energy associated with translational and rotational motion is not preserved. Due to the flexibility of the DSMC method, its adaptation to granular gases is relatively straightforward, even if rotational
motion of grains is taken into account. The present work is a sincere tribute to G. A. Bird's long-lasting influence and accomplishments.

One of the most intriguing phenomena displayed in granular gases (and not present in its \emph{monatomic} molecular gas counterpart, where collisions are elastic) is the absence of energy equipartition among the different degrees of freedom, even in homogeneous and isotropic states.\cite{RN08,G19} In particular, for systems of mechanically different grains (granular mixtures), the mean kinetic translational and rotational energies of each component are in general different.\cite{SKG10,VLSG17,VLSG17b} The lack of energy equipartition is also present in the special cases of one-component rough granular gases\cite{GS95,HZ97,ML98,LHMZ98,CLH02,GNB05,Z06,TWV11,S11a,VT12,KSG14,VSK14,VS15} and smooth granular mixtures.\cite{GD99,WP02,FM02,DHGD02,BT02a,MG02b,BRM05}

It must be noticed that the meaning of \emph{inelastic} collisions used throughout this paper should be distinguished from the one commonly used in polyatomic molecular gases, where collisions make the translational energy to be converted to rotational and vibrational energies or even lead to dissociation or ionization. See, for instance, Refs.\ \onlinecite{A94,VSB99,FY06,YSK07,ZS13,KMK19}

A simple but realistic way of accounting for the effect of inelasticity in the translational and rotational degrees of freedom is by means of a model of inelastic and rough hard spheres. In this model, collisions between spheres of components $i$ and $j$ are characterized by two independent constant coefficients of normal ($\een$) and tangential ($\eet$) restitution.\cite{JR85a,GS95} While the coefficient $0<\een\leq 1$ characterizes the decrease in the magnitude of the normal component of the relative velocity of the points at contact of the colliding spheres, the coefficient $-1\leq \eet\leq 1$ takes into account the change of the tangential component of the relative velocity. Except for $\een=1$ and either $\eet=-1$ (perfectly smooth spheres) or $\eet=1$ (perfectly rough spheres), the total kinetic energy is not conserved in a collision for this model. An interesting feature of the model is that the rotational and translational degrees of freedom of the spheres are coupled through the inelasticity of  collisions.\cite{VSK14}

Since the study of energy nonequipartition in gas mixtures of inelastic rough hard spheres is in general quite complex, it is convenient to consider simple nonequilibrium situations  in order to gain some insight into more general problems. In this paper, we consider the so-called homogeneous cooling state (HCS), namely a spatially uniform state where the (granular) temperature monotonically decays in time.\cite{H83}

As mentioned before, one of the novel features arising from inelasticity is that the \emph{partial} temperatures (measuring the mean kinetic translational and rotational energies of each component) are in general different. More specifically, for a granular mixture one generally has $\Tar\neq \Tat$ for any component $i$, and $T_i^\tr\neq T_j^\tr$ and $T_i^\rot\neq T_j^\rot$ for any pair $i$ and $j$. Here, $T_i^\tr$ and $T_i^\rot$ refer to the translational and rotational temperatures, respectively, of component $i$. The HCS conditions for determining the dependence of the temperature ratios $\Tar/\Tat$, $T_i^\tr/T_j^\tr$, and $T_i^\rot/T_j^\rot$ on the set of coefficients of restitution ($\{\een\}$ and $\{\eet\}$), the concentrations, and the mechanical parameters of the mixture (masses, diameters, and moments of inertia) were obtained in Ref.\ \onlinecite{SKG10} by neglecting (i) correlations between translational and angular velocities and (ii) deviations of the marginal translational velocity distribution from the  Maxwellian. In spite of those approximations, the theoretical results for the temperature ratios have been recently shown\cite{VLSG17,VLSG17b} to present a general good agreement with computer simulations in the tracer limit (a binary mixture where the concentration of one of the components is negligible).

And yet, the fact that  energy equipartition is in  general violated in  granular mixtures does not preclude that, under certain conditions, partial or total equipartition might be present.
To simplify the analysis, we consider here mixtures with common coefficients of restitution ($\alpha$, $\beta$) and reduced moment of inertia ($\q$). Thus, the goal now is to explore whether a particular choice of concentrations, masses, and diameters of the mixture components leads to \emph{partial} energy equipartition, namely $T_i^\tr=T^\tr$ and $T_i^\rot=T^\rot$ (for all $i$), so that the common rotational-to-translational temperature ratio $T^\rot/T^\tr$ coincides with that of a one-component gas of inelastic rough hard spheres.\cite{LHMZ98,Z06} We can think of this phenomenon by imagining that a number of intruder spheres are added to a one-component granular gas and their partial temperatures  \emph{mimic} the corresponding values  of the host gas.\cite{S18,MS19}
Our results show that in fact there are regions in the parameter space of an $s$-component system displaying this \emph{mimicry} effect. More specifically, for given values of the $s-1$ concentration parameters and the $s-1$ diameter ratios, there are $s-1$ conditions whose solution gives the $s-1$ mass ratios such that partial equipartition (in the sense described before) exists.

To assess the accuracy of our approximate theoretical predictions, a comparison with computer simulations has been carried out. In particular, we have numerically solved the Boltzmann kinetic equation via the DSMC method.\cite{B94,B13} In addition, event-driven molecular dynamics (MD) simulations for very dilute systems have also been performed. While the DSMC results assess the reliability of the approximate solution (statistical independence of the translational and angular velocities plus Maxwellian translational distribution), the comparison against MD can be considered as a stringent test of the kinetic equation itself since MD avoids any assumption inherent to kinetic theory (molecular chaos hypothesis). The simulations have been performed in the simple case of a binary mixture ($s=2$) where one of the components (say $i=1$) is present in tracer concentration (i.e., $n_1/n_2\to 0$, $n_i$ being the number density of component $i$). This problem is equivalent to that of an impurity or intruder immersed in a granular gas of rough spheres (component $2$).\cite{VGK14,S18b} This implies that (a) the state of the excess component is not perturbed by the presence of the tracer particles (so that its velocity distribution function $f_2$ obeys the closed Boltzmann equation for a one-component granular gas) and, additionally, (b) collisions among tracer particles can be neglected in the kinetic equation for the distribution function $f_1$ (Boltzmann--Lorentz equation). In
this limiting case,  the three relevant temperature ratios (namely $T_1^\tr/T_2^\tr$, $T_1^\rot/T_1^\tr$, and $T_2^\rot/T_2^\tr$) are in general functions of  $\alpha$, $\beta$,  $\kappa$,  the mass ratio $m_1/m_2$, and the diameter ratio $\sigma_1/\sigma_2$. As we will see, the  conditions for mimicry (i.e., $T_1^\tr/T_2^\tr=1$, $T_1^\rot/T_1^\tr=T_2^\rot/T_2^\tr$) stemming from our approximation turn out to be independent of $\alpha$, $\beta$, and $\q$.

The paper is organized as follows.  The kinetic theory for multicomponent granular gases is briefly summarized in Sec.\ \ref{sec2}. Section \ref{sec3} deals with the explicit determination of the so-called production rates when the marginal translational distribution is approximated by a Maxwellian distribution. Starting from these general expressions, the conditions for the mimicry effect are obtained in Sec.\ \ref{sec4} for an $s$-component mixture and, next, particularized to a binary mixture ($s=2$). Section \ref{sec5} focuses on the comparison between the approximate results and computer simulations performed in the tracer limit for some representative systems. The paper is closed in Sec.\ \ref{sec6} with a brief discussion of the main results reported here.

\section{Boltzmann equation for granular mixtures of rough spheres}
\label{sec2}

We consider an $s$-component gas of inelastic rough hard spheres. Particles of component $i$ have a mass $\ma$, a diameter $\da$, and a
moment of inertia $\Ia=\qa \ma\da^2/4$. The reduced moment of inertia $\qa$
ranges from $\qa=0$ (mass concentrated in the center) to $\qa=\frac{2}{3}$
(mass concentrated in the surface). If the mass of a particle of component $i$ is uniformly distributed, then $\qa=\frac{2}{5}$.
The inelasticity and roughness of colliding particles are characterized by the set of coefficients of normal ($\alpha_{ij}$) and tangential ($\beta_{ij}$) restitution. Those coefficients of restitution are defined by the collision rule
\begin{equation}
\kk\cdot \wwwab'=-\een \kk\cdot \wwwab,\quad \kk\x \wwwab'=-\eet \kk\x \wwwab,
\label{restitution}
\end{equation}
where $\wwwab$ and $\wwwab'$ are the pre- and post-collisional relative velocities of the points at contact of two colliding spheres of components $i$ and $j$, and  $\kk$ is the unit vector joining their centers.
As said before, while the coefficient $\een$ ranges from $\een=0$ (perfectly inelastic particles) to $\een=1$ (perfectly elastic particles),  the coefficient $\eet$ runs from $\eet=-1$ (perfectly smooth particles) to $\eet=1$ (perfectly rough particles). Except if $\een=1$ and $|\eet|=1$, kinetic energy is dissipated upon a collision $ij$.

At a kinetic level, all the relevant information is contained in the velocity distribution function $f_i(\mathbf{v},\ww;t)$ of each component, where we have particularized to homogeneous states. Here, $\cc$ and $\ww$ denote the translational and angular velocities, respectively.
From the knowledge of $f_i$ one can obtain the number density and the so-called translational and rotational (partial) temperatures of component $i$  as
\begin{subequations}
\label{III.3}
\begin{equation}
 \na=\int d\cc \int d\ww\,  \fa(\cc,\ww;t),
\end{equation}
\begin{equation}
 \Tat(t)=\frac{\ma}{3\na}\int d\cc \int d\ww\, v^2 \fa(\cc,\ww;t),
\end{equation}
\begin{equation}
 \Tar(t)=\frac{\Ia}{3\na}\int d\cc \int d\ww\, \omega^2 \fa(\cc,\ww;t).
\end{equation}
\end{subequations}
As a measure of the total kinetic energy per particle, one can define the \emph{total} temperature as
\begin{equation}
T=\sum_{\al=1}^s\frac{\na}{2n}\left(\Tat+\Tar\right),
\label{III.13}
\end{equation}
where $n=\sum_{\al=1}^s \na$ is the total number density.

In the low density regime ($n_i\da^3\ll 1$),  the  velocity distribution functions obey a closed set of coupled Boltzmann equations,\cite{RN08,SKG10,G19}
\begin{equation}
\partial_t \fa(\cc,\ww;t)=\sum_{\be=1}^s \Qab[\cc,\ww;t|\fa,\fb],
\label{2}
\end{equation}
where
\bal
\Qab&[\cc_1,\ww_1;t|\fa,\fb]=\dab^2\int d\cc_2\int d\ww_2\int d\kk\,\nn &\times\Theta(\cc_{12}\cdot\kk)(\cc_{12}\cdot\kk)\Bigg[\frac{1}{\een^2\eet^2}\fa(\cc_1'',\ww_1'';t)\fb(\cc_2'',\ww_2'';t)\nn
&-
\fa(\cc_1,\ww_1;t)\fb(\cc_2,\ww_2;t)\Bigg]
\label{III.2}
\eal
is the collision operator. Here,  $\Theta(x)$ is the Heaviside step function, $\dab\equiv (\da+\db)/2$, $\cc_{12}=\cc_1-\cc_2$ is the relative translational velocity, and the double primes denote precollisional velocities. Note that Eq.\ \eqref{2} describes a freely cooling (or undriven) granular gas, so that the total kinetic energy decays monotonically in time.\cite{H83,BP04,G19}
The evolution equations for the partial translational ($\Tat$) and rotational ($\Tar$) temperatures can be obtained by multiplying  Eq.\ \eqref{2} by $\ma v^2$ and $\Ia \omega^2$, respectively, and integrating over the velocities. The results are
\begin{equation}
\label{evol_T}
\partial_t \Tat=-\zt_i \Tat,\quad \partial_t \Tar=- \zr_i \Tar,
\end{equation}
with
\begin{equation}
\zt_i=\sum_{j=1}^s \zabt ,\quad \zr_i=\sum_{j=1}^s \zabr.
\end{equation}
Here,
\begin{subequations}
\label{54}
\begin{equation}
\zabt\equiv -\frac{\ma}{3\na\Tat}\int d\cc\int d\ww\, v^2 \Qab[\cc,\ww;t|\fa,\fb],
\label{54a}
\end{equation}
\begin{equation}
\zabr\equiv -\frac{\Ia}{3\na\Tar}\int d\cc\int d\ww\, \omega^2 \Qab[\cc,\ww;t|\fa,\fb]
\label{54b}
\end{equation}
\end{subequations}
are energy production rates.

Binary collisions produce two main effects.\cite{S11b,MS18} On the one hand,  a certain energy transfer exists between the two components involved and also from rotational to translational (or vice versa) kinetic energy. On the other hand, part of the total kinetic energy of both particles is dissipated and goes to increase the internal agitation of the molecules the grains are made of. Thus, as said before, in the absence of any external driving, the total granular temperature $T$ monotonically decays with time (Haff's law\cite{H83}).
However, after a certain transient stage lasting typically less than $100$ collisions per particle,\cite{VSK14} a scaling regime is reached (the so-called HCS) such that all the time dependence of the distributions $f_i$  occurs through $T$.\cite{D00,VSK14} This  implies $\partial_t
( \Tat/T )=\partial_t ( \Tar/T )=0$ for all $i$, and hence
\begin{equation}
\label{HCS}
\zt_1=\cdots=\zt_s=\zr_1=\cdots=\zr_s.
\end{equation}

It must be stressed that the energy production rates $\zabt$ and $\zabr$ are in general complex functionals of the distribution functions $\fa$ and $\fb$, so that the set of Eqs.\ \eqref{evol_T} is not closed and Eq.\ \eqref{HCS} cannot be solved exactly, unless an approximate closure is assumed.

\section{Maxwellian approximation}
\label{sec3}

In order to determine the production rates $\zabt$ and $\zabr$ in terms of the partial temperatures $\Tat$, $\Tar$, $\Tbt$, and $\Tbr$, we assume that the velocity distributions in Eqs.\  \eqref{54} can be approximated by considering that (i) the translational and rotational velocities are statistically independent and (ii) the marginal translational distribution  is a Maxwellian function. More specifically,
\begin{equation}
\label{IV.1}
\fa(\cc,\ww)\to  \left(\frac{\ma}{2\pi\Tat}\right)^{3/2}\exp\left(-\frac{\ma v^2}{2\Tat}\right)
\far(\ww),
\end{equation}
where $\far(\ww)$ is the marginal rotational distribution function.
By inserting Eq.\ \eqref{IV.1} into Eqs.\ \eqref{54}, and after some algebra, one obtains the explicit expressions\cite{SKG10,S11b,MS18}
\begin{widetext}
\begin{subequations}
\label{55}
\begin{equation}
\zabt=\frac{2\mab^2\nu_{\al\be}}{3\ma\Tat}
\left[\left({\en+\et}\right)\frac{2\Tat}{\mab}-
\left({\en^2+\et^2}\right)\left(\frac{\Tat}{\ma}+\frac{\Tbt}{\mb}\right)
-{\et^2}\left(\frac{\Tar}{\ma\qa}+\frac{\Tbr}{\mb\qb}\right)
\right],
\label{55a}
\end{equation}
\begin{equation}
\zabr=\frac{2\mab^2\nu_{\al\be}\et}{3\ma\qa\Tar}\left[\frac{2\Tar}{\mab}-{\et}\left(\frac{\Tat}{\ma}+\frac{\Tbt}{\mb}+\frac{\Tar}{\ma\qa}+\frac{\Tbr}{\mb\qb}\right)
\right],
\label{56}
\end{equation}
\end{subequations}
\end{widetext}
where we have introduced
the quantities
\begin{subequations}
\label{20}
\begin{equation}
\label{20a}
\en\equiv1+\een,\quad\et\equiv\frac{\qab}{1+\qab}\left(1+\eet\right),
\end{equation}
\begin{equation}
\label{20b}
\mab\equiv \frac{\ma\mb}{\ma+\mb},\quad \qab\equiv \qa\qb\frac{\ma+\mb}{\qa\ma+\qb\mb},
\end{equation}
\end{subequations}
and the effective collision frequencies
\begin{equation}
\nu_{\al\be}\equiv2\sqrt{2\pi}\nb{\dab^2}\sqrt{\frac{\Tat}{\ma}+\frac{\Tbt}{\mb}}.
\label{56b}
\end{equation}
In summary, in an $s$-component mixture, insertion of Eqs.\ \eqref{55}  into Eq.\ \eqref{HCS} provides a closed set of $2s-1$  coupled algebraic equations for the $2s-1$ independent temperature ratios, which in general must be solved numerically.

\begin{figure}
\includegraphics[width=0.9\columnwidth]{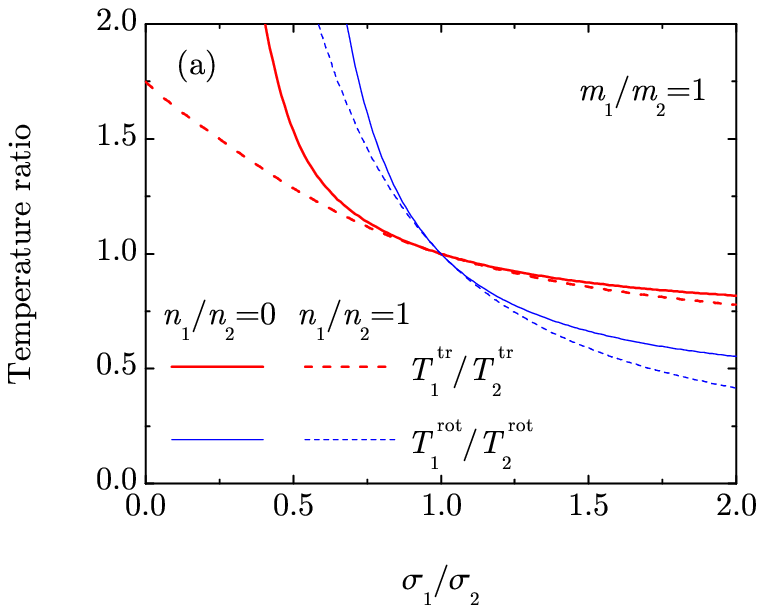}\\
\includegraphics[width=0.9\columnwidth]{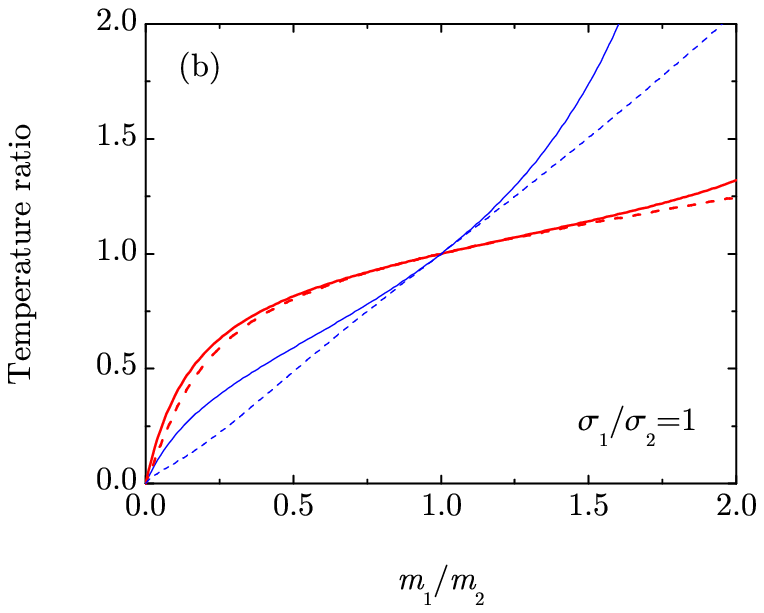}
\caption{Plot of the temperature ratios $\Tt_1/\Tt_2$ (thick red lines) and $\Tr_1/\Tr_2$  (thin blue lines) for the extreme concentrations $n_1/n_2=0$ (solid lines) and $n_1/n_2=1$ (dashed lines). In panel (a) the temperature ratios are plotted vs the size ratio $\sigma_1/\sigma_2$ at equal masses ($m_1/m_2=1$), while in panel (b) the temperature ratios are plotted vs the mass ratio $m_1/m_2$ at equal sizes ($\sigma_1/\sigma_2=1$). In both cases, $\alpha=0.8$ and $\beta=0$.}
\label{fig1}
\end{figure}

It must be remarked that, due to the inelastic character of collisions, the HCS distribution is known to deviate from the simple form \eqref{IV.1}. First, the translational distribution function presents fat high-velocity tails\cite{EP97,vNE98} and the velocity cumulants are not negligible;\cite{GD99b,MS00,SM09,SKS11,VSK14} second, statistical correlations  between translational and angular velocities  have been predicted theoretically and confirmed by simulations.\cite{BPKZ07,KBPZ09,SKS11,VS15,VSK14} However, the impact of those limitations of Eq.\ \eqref{IV.1} on the evaluation of the granular temperatures is not strong.\cite{VSK14} Thus, despite the simplicity of the approximation \eqref{IV.1}, numerical results obtained from the DSMC and MD methods for a binary mixture with a tracer component compare very well with the theoretical results derived from Eqs.\ \eqref{55}.\cite{VLSG17b}

The parameter space in an $s$-component mixture is made of $s^2+5s-3$ quantities: $s(s+1)/2$ coefficients of normal restitution, $s(s+1)/2$ coefficients of tangential restitution, $s-1$ mole fractions, $s-1$ mass ratios, $s-1$ diameter ratios, and $s$ reduced moments of inertia. To illustrate the impact of both mass and size on the temperature ratios, let us consider a binary mixture ($s=2$) with common coefficients of restitution and common reduced moments of inertia (i.e., $\alpha_{ij}=\alpha$, $\beta_{ij}=\beta$, and $\qa=\q$). Without loss of generality we assume that $n_1\leq n_2$. Otherwise, the size ratio $\sigma_1/\sigma_2$ and the mass ratio $m_1/m_2$ are arbitrary.

Let us first suppose that $m_1/m_2=1$ and qualitatively analyze the influence of the diameter ratio $\sigma_1/\sigma_2$ on the component-component temperature ratios $\Tt_1/\Tt_2$ and $\Tr_1/\Tr_2$. According to Eq.\ \eqref{56b}, if $\sigma_1/\sigma_2<1$  component $1$ collides less frequently than component $2$ and hence it dissipates less kinetic energy. Therefore, one may expect $\Tt_1/\Tt_2>1$ and $\Tr_1/\Tr_2>1$. The opposite can be expected if $\sigma_1/\sigma_2>1$. This qualitative analysis is confirmed by Fig.\ \ref{fig1}(a) for a representative case, where it can be observed that the temperature ratios $\Tt_1/\Tt_2$ and $\Tr_1/\Tr_2$ monotonically decrease with increasing $\sigma_1/\sigma_2$, regardless of the value of the concentration.
As for the influence of the mass ratio (assuming now $\sigma_1/\sigma_2=1$), it is less straightforward than the influence of the size ratio. If initially all the temperatures are equal, Eqs.\ \eqref{55} and \eqref{56b} show that the more massive particles have a smaller cooling rate. As a consequence, once the asymptotic HCS is reached, one expects the more massive spheres to have a larger temperature. This is confirmed by Fig.\ \ref{fig1}(b), which shows a monotonic increase of both $\Tt_1/\Tt_2$ and $\Tr_1/\Tr_2$  with increasing $m_1/m_2$, again with independence of the concentration.

Typically, the bigger spheres are also the heavier ones and, therefore, whether the ratios $\Tt_1/\Tt_2$ and $\Tr_1/\Tr_2$ are smaller or larger than unity results from the competition between both mechanisms exemplified by Fig.\ \ref{fig1}. Thus, it might be possible that a certain coupling between  $\sigma_1/\sigma_2$ and $m_1/m_2$ leads to $\Tt_1/\Tt_2=1$ and $\Tr_1/\Tr_2=1$. As mentioned in Sec.\ \ref{sec1}, this is what we refer to as the mimicry effect.

\section{Mimicry effect}
\label{sec4}
Let us consider again an $s$-component mixture particularized to the case of equal coefficients of restitution and reduced moments of inertia, i.e., $\alpha_{ij}=\alpha$, $\beta_{ij}=\beta$, and $\kappa_i=\kappa$.
The question we want to address is under which conditions the mixture exhibits \emph{partial} equipartition in the sense that $\Tt_i=\Tt$ and $\Tr_i=\Tr$, even though $\Tt\neq\Tr$, the ratio $\Tr/\Tt$ being the same as that of a one-component granular gas.\cite{LHMZ98,Z06} If that is the case, we can say that the mixture \emph{mimics} a one-component gas in the above sense.

By setting $\Tt_i=\Tt$ and $\Tr_i=\Tr$ in Eqs.\ \eqref{55}, one obtains
\begin{subequations}
\label{Xij-F}
\begin{equation}
\zabt=X_{ij}\sqrt{\Tt}F^\tr(\theta),
\end{equation}
\begin{equation}
\zabr=X_{ij}\sqrt{\Tt}F^\rot(\theta),
\end{equation}
\end{subequations}
where $\theta\equiv \Tr/\Tt$ and
\begin{subequations}
\begin{equation}
X_{ij}=\frac{4\sqrt{2\pi}}{3}\frac{\sqrt{\mab} \nb\dab^2}{\ma},
\end{equation}
\begin{equation}
F^\tr(\theta)=\overline{\alpha}(2-\overline{\alpha})+\overline{\beta}(2-\overline{\beta})-\frac{\overline{\beta}^2}{\q}\theta,
\end{equation}
\begin{equation}
F^\rot(\theta)=\frac{\overline{\beta}}{\q^2\theta}\left[(2\kappa-\overline{\beta})\theta-\overline{\beta}\q\right].
\end{equation}
\end{subequations}
Here, according to Eqs.\ \eqref{20}, $\overline{\alpha}=1+\alpha$ and $\overline{\beta}=(1+\beta)\q/(1+\q)$.
It is noteworthy that in the factorizations \eqref{Xij-F} the quantity $X_{ij}$  is the same in $\zabt$ and $\zabr$, and depends only on the concentrations, masses, and diameters of the spheres. In contrast, the functions $F^\tr(\theta)$ and $F^\rot(\theta)$ only depend on the temperature ratio $\theta$ and the mechanical parameters $\alpha$, $\beta$, and $\q$.

The HCS conditions \eqref{HCS} decouple into
\begin{equation}
\label{Xs}
X_1=\cdots=X_s,\quad X_i\equiv \sum_{j=1}^s X_{ij},
\end{equation}
and
$F^\tr(\theta)=F^\rot(\theta)$. From the latter equality, one easily gets
\begin{subequations}
\label{thetah}
\begin{equation}
\theta=\sqrt{1+h^2}+h,
\end{equation}
\begin{equation}
h=\frac{(1+\q)^2}{2\q(1+\beta)^2}\left[1-\alpha^2-(1-\beta^2)\frac{1-\q}{1+\q}\right].
\end{equation}
\end{subequations}
For a general $s$-component mixture, Eq.\ \eqref{Xs} gives  $s-1$ conditions for the $3(s-1)$ density, mass, and diameter ratios.
In the particular case of a binary mixture ($s=2$), the single condition on $n_1/n_2$, $m_1/m_2$, and $\sigma_1/\sigma_2$ is
\begin{equation}
\frac{n_1}{n_2}=\frac{(1+\frac{\sigma_1}{\sigma_2})^2\sqrt{\frac{m_2}{m_1}}-4\sqrt{\frac{m_1+m_2}{2m_2}}}{(1+\frac{\sigma_1}{\sigma_2})^2\sqrt{\frac{m_1}{m_2}}
-4\frac{\sigma_1^2}{\sigma_2^2}\sqrt{\frac{m_1+m_2}{2m_1}}}.
\label{115}
\end{equation}
Equation \eqref{115} is equivalent to a quadratic equation for $\sigma_1/\sigma_2$ and a quartic equation for $m_1/m_2$.
In the special tracer limit ($n_1/n_2\to 0$), the quartic equation for the mass ratio reduces to a quadratic equation whose solution is
\begin{equation}
\label{tracer_limit}
\frac{m_1}{m_2}=\frac{1}{2}\left[\sqrt{1+\frac{1}{2}\left(1+\frac{\sigma_1}{\sigma_2}\right)^4}-1\right].
\end{equation}
In this tracer limit, $m_1/m_2$ has a lower bound $\mu_-(0)=(\sqrt{3/2}-1)/2\simeq 0.11$ (corresponding to $\sigma_1/\sigma_2\to 0$) but it does not have any upper bound. However, for finite concentration ($n_1/n_2\neq 0$), an additional finite upper bound (corresponding to $\sigma_1/\sigma_2\to\infty$) exists, namely  $\mu_-(n_1/n_2)\leq m_1/m_2\leq \mu_+(n_1/n_2)$, where
\begin{equation}
\mu_-(x)=\frac{2\sqrt{2}\sqrt{3+x}-x-4}{8-x^2},\quad \mu_+(x)=\frac{1}{\mu_-(x^{-1})}.
\end{equation}

\begin{figure}
\includegraphics[width=0.9\columnwidth]{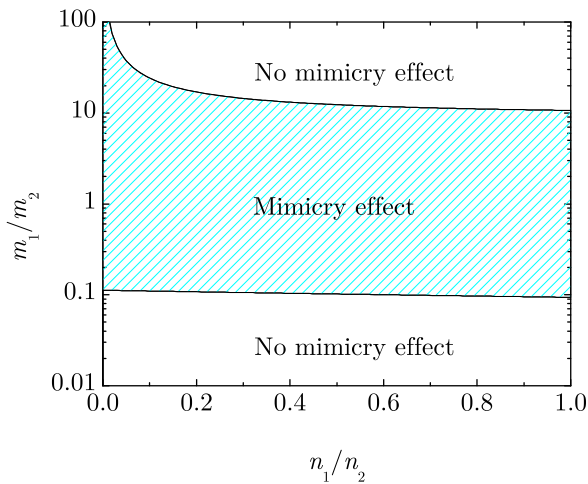}
\caption{Plot of $\mu_-(n_1/n_2)$ (lower curve) and $\mu_+(n_1/n_2)$ (upper curve) vs the concentration parameter $n_1/n_2$. The mimicry effect is possible only in the region $\mu_-(n_1/n_2)\leq m_1/m_2\leq \mu_+(n_1/n_2)$.}
\label{fig3}
\end{figure}

The dependence of $\mu_-(n_1/n_2)$ and $\mu_+(n_1/n_2)$ on the concentration parameter $n_1/n_2$ is shown in Fig.\ \ref{fig3}. We observe that $\mu_-(n_1/n_2)$ presents a very weak dependence on the concentration. On the other hand, $\mu_+(n_1/n_2)$ increases rapidly as one approaches the tracer limit, diverging at $n_1/n_2=0$. The mimicry effect is possible only inside the shaded region of Fig.\ \ref{fig3}; outside of that region, Eq.\ \eqref{115} fails to provide physical solutions for $\sigma_1/\sigma_2$.

\begin{figure}
\includegraphics[width=0.9\columnwidth]{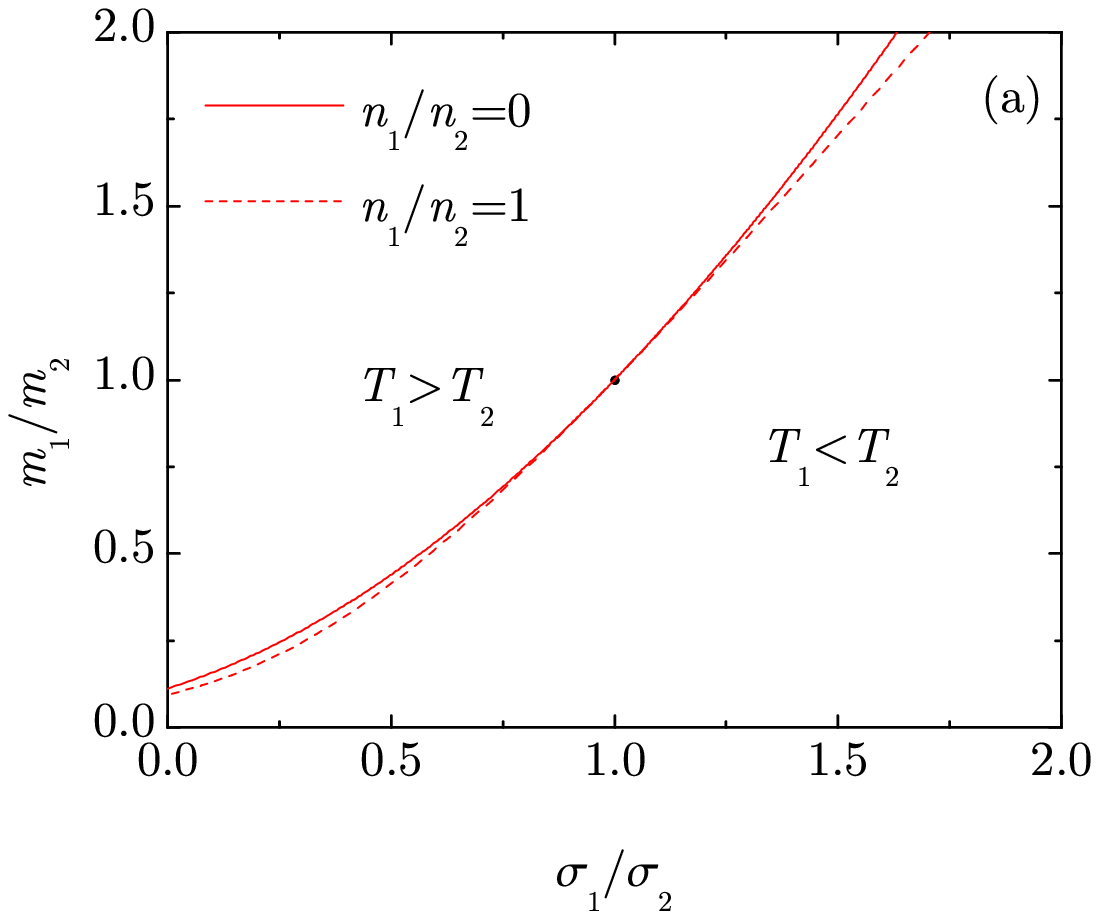}\\
\includegraphics[width=0.9\columnwidth]{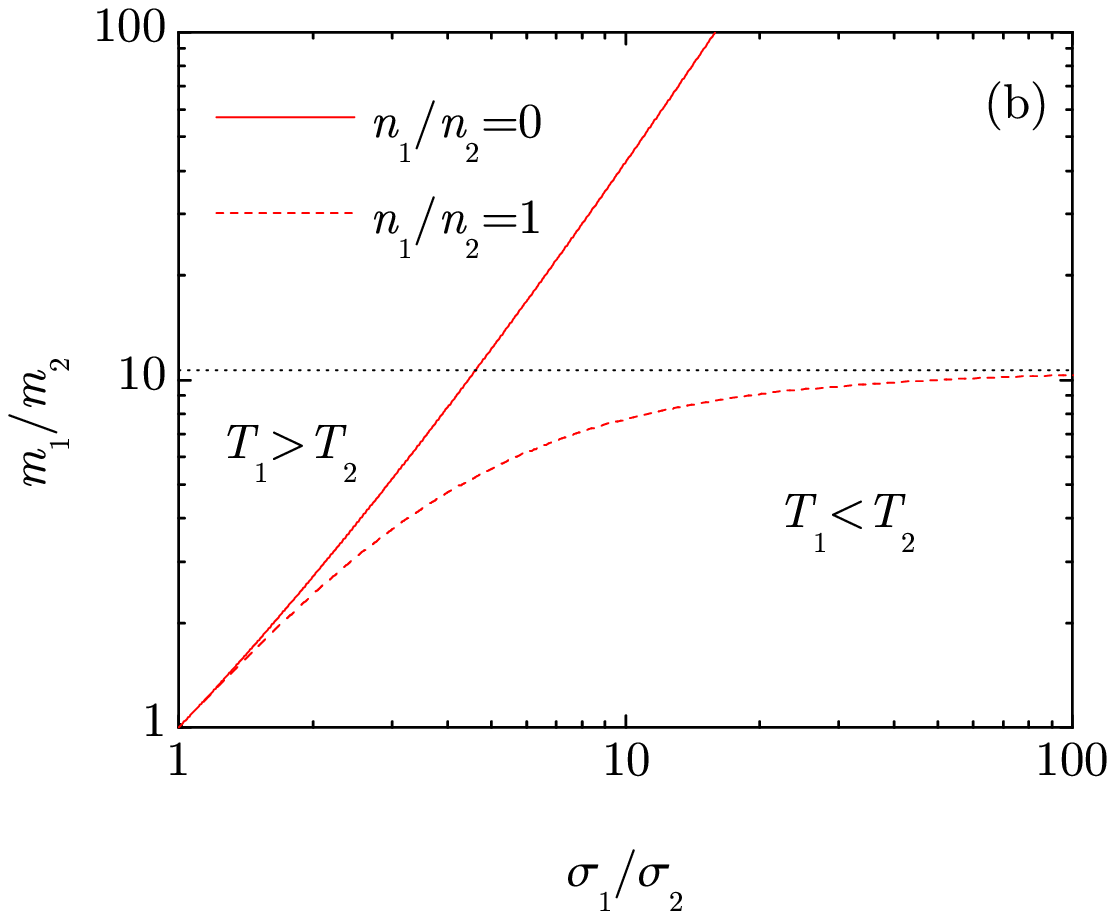}
\caption{Phase diagram in the plane $m_1/m_2$ vs $\sigma_1/\sigma_2$. The solid (dashed) curve represents the locus of mimicry in the tracer (equimolar) limit $n_1/n_2\to 0$ ($n_1/n_2\to 1$). Panel (a) shows the region $0\leq\sigma_1/\sigma_2\leq 2$, $0\leq m_1/m_2\leq 2$, while panel (b) shows (in logarithmic scale) the region $1\leq\sigma_1/\sigma_2\leq 100$, $1\leq m_1/m_2\leq 100$. In panel (b), the horizontal dotted line is the asymptote $m_1/m_2\to \mu_+(1)\simeq 10.657$ corresponding to the limit $\sigma_1/\sigma_2\to \infty$ in the equimolar case.}
\label{fig2}
\end{figure}

Figure \ref{fig2} plots the values of $m_1/m_2$ vs $\sigma_1/\sigma_2$ exhibiting the mimicry effect for two extreme concentrations, i.e., $n_1/n_2\to 0$ (tracer limit) and $n_1/n_2=1$ (equimolar mixture). The curves corresponding to intermediate concentrations lie in the region between those two lines.  While a weak influence of the concentration can be observed if $\sigma_1/\sigma_2\lesssim 2$, the influence becomes very strong if $\sigma_1/\sigma_2$ is very large. In particular, in the limit $\sigma_1/\sigma_2\to\infty$, the mass ratio $m_1/m_2$ tends to its asymptotic value $\mu_+(1)\simeq 10.657$  in the equimolar case, but it diverges as $m_1/m_2\approx (\sigma_1/\sigma_2)^2/2\sqrt{2}$ in the tracer limit.

Given a value of $n_1/n_2$, the locus $m_1/m_2$ vs $\sigma_1/\sigma_2$ splits the plane into two regions. In the points below the locus curve, $T_1^\tr<T_2^\tr$ and $T_1^\rot<T_2^\rot$ as a consequence of the competition between the size and mass effects  previously discussed in connection with Fig.\ \ref{fig1}. Alternatively, $T_1^\tr>T_2^\tr$ and $T_1^\rot>T_2^\rot$ in the points above the locus curve.
It is interesting to note that the curve  representing equal  particle mass density, $m_1/m_2=(\sigma_1/\sigma_2)^3$, lies below the mimicry curve if $\sigma_1<\sigma_2$ and above it if $\sigma_1>\sigma_2$. Thus, if the mass density of both types of spheres is the same, the bigger spheres have a larger (translational or rotational) temperature than the smaller spheres. On the other hand, the mimicry effect requires the bigger spheres to be less dense than the smaller spheres.

\begin{figure}
\includegraphics[width=\columnwidth]{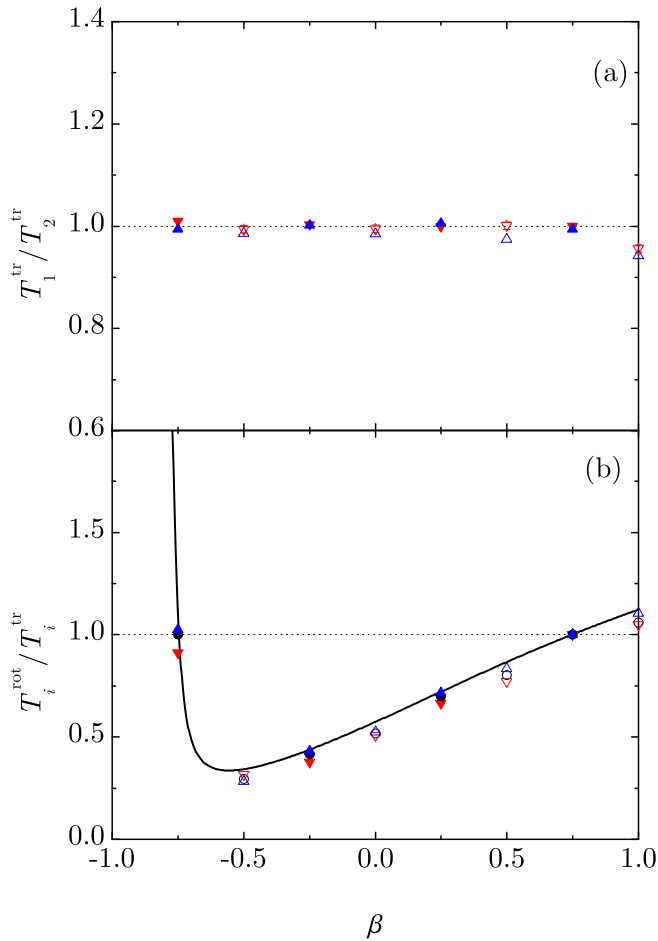}
\caption{Plot of the simulation results for (a) $\Tt_1/\Tt_2$ and (b) $\Tar/\Tat$ ($i=1,2$) vs the coefficient of tangential restitution $\beta$  at $\alpha=0.9$ and $\q=\frac{2}{5}$. In panels (a) and (b) the (red) down triangles correspond to a small intruder with $\sigma_1/\sigma_2=\frac{1}{2}$ and $m_1/m_2=0.440$, while the (blue) up triangles correspond to a big intruder with $\sigma_1/\sigma_2={2}$ and $m_1/m_2=2.721$. In panel (b) the (black) circles represent the  temperature ratio $\Tr_2/\Tt_2$ of the host gas, while the solid line is the one-component theoretical prediction. The filled symbols at $\beta=-0.75, -0.25, 0.25, 0.75$ represent DSMC data and the open symbols at $\beta=-0.5, 0, 0.5, 1$ represent MD data. The error bars are of similar size as the symbols.}
\label{fig4}
\end{figure}

Since the mimicry conditions \eqref{Xs} are independent of the values of $\alpha$, $\beta$, and $\q$, they are the same conditions as for equipartition in the smooth-sphere case ($\beta=-1$), where only the translational temperatures are relevant. In the rough-sphere case, however, complete equipartition is  not fulfilled since $\Tt$ and $\Tr$ are, in general, different. Full energy equipartition (i.e., $\Tt=\Tr$) is achieved if $F^\tr(1)=F^\rot(1)$, i.e., $\alpha^2(1+\q)=\beta^2(1-\q)+2\q$.

\section{Comparison with computer simulations}
\label{sec5}
The theoretical predictions discussed in Sec.\ \ref{sec4} for the mimicry effect are based on the simple ansatz \eqref{IV.1}. On the other hand, previous results\cite{BPKZ07,SKS11,VSK14} show that statistical correlations between the translational and angular velocities, as well as cumulants of the translational distribution, can be observed. Therefore, it is important to assess the reliability of the theoretical results based on Eq.\ \eqref{IV.1} by comparison with computer simulations.

For the sake of simplicity, we consider here an intruder (component $1$) immersed in a one-component granular gas (component $2$). This is equivalent to a binary mixture in the tracer limit ($n_1/n_2\to 0$). In addition, $\alpha_{12}=\alpha_{22}=\alpha$, $\beta_{12}=\beta_{22}=\beta$, and $\q_1=\q_2=\q$. Two representative cases are studied: a small intruder ($\sigma_1/\sigma_2=\frac{1}{2}$) and a big intruder ($\sigma_1/\sigma_2=2$). The masses of the intruders are taken  as the values  for which, according to Eq.\ \eqref{tracer_limit}, a mimicry effect is expected. More specifically, $m_1/m_2=\sqrt{\frac{113}{28}}-\frac{1}{2}\simeq0.440$ and $m_1/m_2=\sqrt{\frac{83}{8}}-\frac{1}{2}\simeq 2.721$ for $\sigma_1/\sigma_2=\frac{1}{2}$ and $\sigma_1/\sigma_2=2$, respectively. Thus, the small  intruder is $(m_1/\sigma_1^3)/(m_2/\sigma_2^3)=3.52$ times \emph{denser} than a particle of the host gas, while the big intruder is $(m_2/\sigma_2^3)/(m_1/\sigma_1^3)=2.94$ times \emph{less dense} than a particle of the host gas.

In the simulations, the  values of the coefficient of tangential restitution are $\beta=-0.75, -0.25, 0.25, 0.75$ (DSMC) and $\beta=-0.5, 0, 0.5, 1$ (MD), while the coefficient of normal restitution in both sorts of simulation is chosen as $\alpha=0.9$. Figure \ref{fig4} displays the simulation values of the three independent temperature ratios $\Tt_1/\Tt_2$ [panel (a)], $\Tr_1/\Tt_1$, and $\Tr_2/\Tt_2$ [panel (b)].
One can observe from Fig.\ \ref{fig4}(a) that the translational temperatures of both the small and big intruders are indeed very close to that of the host gas. The larger deviation of $\Tt_1/\Tt_2$ from unity (about $5\%$) appears at $\beta=1$, but even in that case $\Tt_1$ is practically the same for the small and big intruders.
As a complement, Fig.\ \ref{fig4}(b) exhibits a rather good collapse of the rotational-to-translational temperature ratio for the small and big intruders and the host gas. On the other hand, the simulation data show a  tendency of that ratio to be slightly smaller (larger) for the small (big) intruder than for the host gas. Apart from that, the theoretical prediction \eqref{thetah} for $\Tr_i/\Tt_i$ agrees well with the simulation results, although it tends to overestimate them. In summary, Fig.\ \ref{fig4} shows that the simulation data confirm reasonably well the theoretical prediction of the mimicry effect. Interestingly, at $|\beta|= 0.75$, the DSMC results exhibit a high degree of full equipartition, in agreement with the theoretical expectation at $|\beta|=\sqrt{\frac{167}{300}}\simeq 0.746$.

\section{Discussion}
\label{sec6}

It is well known that in a multicomponent gas of inelastic and rough hard spheres the total kinetic energy is not equally partitioned among the different degrees of freedom. This implies that the translational and rotational temperatures associated with each component are in general different.

It is of physical interest to find regions of the system's parameter space where a certain degree of energy equipartition shows up (effect that we denote as \textit{mimicry}). Here, we have focused on the HCS of systems with common values of the coefficients of normal and tangential restitution (i.e., $\alpha_{ij}=\alpha$, $\beta_{ij}=\beta$), as well as of the reduced moment of inertia (i.e., $\qa=\q$), and have addressed the question of whether all the components of the mixture mimic a one-component system in the sense that they adopt the same rotational and translational temperatures as the latter.

From a simple approximation, where (i) the statistical correlations between the translational and angular velocities are neglected and (ii) the  marginal translational distribution function is approached by a Maxwellian, we have determined the conditions \eqref{Xs} for the mimicry effect. Interestingly, those approximate conditions are ``universal'' in the sense that they are independent of the values of $\alpha$, $\beta$, and $\kappa$. In fact, they are the same conditions as for equipartition in the case of smooth spheres ($\beta=-1$). For a mixture with an arbitrary number of components, and given the mole fractions and the diameter ratios, those conditions provide the mass ratios for which mimicry is present. In the particular case of a binary mixture, there is a single condition given by Eq.\ \eqref{115}. As can be seen from Fig.\ \ref{fig3}, the mass ratio has lower and upper bounds, which depend on the concentration.
This means that if the mass ratio is outside of the above interval, no mimicry effect is possible, no matter the value of the size ratio.

To assess the theoretical predictions, computer simulations have been carried out in the tracer (or impurity) limit, where the mimicry condition becomes quite simple, as can be seen from Eq.\  \eqref{tracer_limit}. Both DSMC and MD simulations present a good agreement with the theoretical results, as shown in Fig.\ \ref{fig4}. While the DSMC results gauge the reliability of the assumptions (i) and (ii) described in the preceding paragraph, the MD results go beyond that, since they are free from the molecular chaos assumption. Therefore, the agreement between the kinetic theory approximations and the MD data can be considered as a relevant result of the present paper.

The simplicity of the theoretical analysis for mimicry carried out in this work is heavily based on the assumption that the coefficients of normal and tangential restitution and the reduced moment of inertia of the impurity are the same as those of the particles of the host gas. This seems to be at odds with the fact that the mimicry effect requires the smaller spheres to have a higher particle mass density than the bigger spheres. A way of circumventing this problem is by tailoring the impurity particles with a nonuniform mass distribution made of three concentric shells, so that the external shell is made of the same material as that of the host particles. This is worked out in the Appendix.

While in this paper our focus has been mainly academic and driven by purely scientific interest, the mimicry effect discussed here might find some practical applications. For instance, the translational  and rotational temperatures of a granular gas could be probed by introducing in the gas a few  bigger tracer particles with appropriate particle mass densities.

\begin{acknowledgments}
This paper was dedicated to the memory of Mar\'ia Jos\'e Ruiz-Montero, who pioneered the application of Bird's DSMC method to granular gases.
This work has been supported by the Spanish  Agencia Estatal de
Investigaci\'on Grants (partially financed by the ERDF)
Nos.~MTM2017-84446-C2-2-R (A.L.) and FIS2016-76359-P
  (F.V.R., V.G, and A.S.),  and by the
  Junta de Extremadura (Spain) Grants Nos.~IB16013, IB16087, and GR18079,
  partially funded by the ERDF, (F.V.R., V.G., and A.S.).  A.L. thanks the hospitality of the SPhinX group of the University of Extremadura, where part of this work was done.
  Use of
  computing facilities from Extremadura Research Center for Advanced
  Technologies (CETA-CIEMAT), funded by the ERDF, is also
  acknowledged.
\end{acknowledgments}

\appendix*

\begin{figure}
\includegraphics[width=0.8\columnwidth]{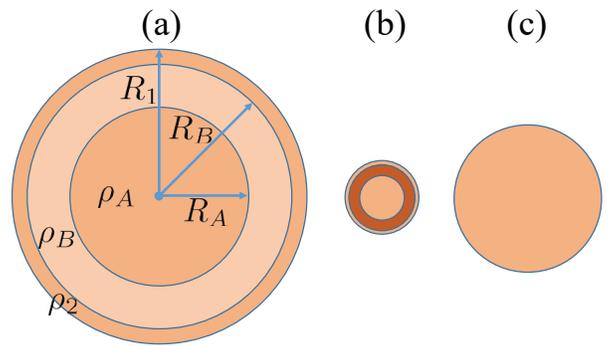}
\caption{(a) Sketch of a sphere with a nonuniform mass distribution of average density $\rho_1$ and a reduced moment of inertia $\q=\frac{2}{5}$. In this case, $\rho_B<\rho_1<\rho_A=\rho_2$, so that $R_A/R_1\simeq0.61$ and $R_B/R_1\simeq0.90$. (b) The same, except that $\rho_B>\rho_1>\rho_A=\rho_2$. (c) Sphere with a uniform mass distribution of density $\rho_2$ and reduced moment of inertia $\q=\frac{2}{5}$. Panels (a) and (b) represent intruders with $\sigma_1/\sigma_2=2$ and $\frac{1}{2}$, respectively, while panel (c) represents a particle of the host gas. If (a) $m_1/m_2= 2.721$  and (b) $m_1/m_2= 0.440$, then $\rho_B/\rho_2=0.32$ and $6.04$, respectively.}
\label{fig5}
\end{figure}

\section{Tailoring the moment of inertia and mass of a sphere}
Consider a sphere of mass $m_1$ and radius $R_1=\frac{1}{2}\sigma_1$ with a nonuniform radial distribution of mass. More specifically, we assume that the sphere is made of an inner core of density $\rho_A$ and radius $R_A$, a spherical shell of density $\rho_B$ and radii $R_A$ and $R_B$, and finally an outer spherical shell of density $\rho_2$ and radii $R_B$ and $R_1$ [see Fig.\ \ref{fig5}(a)]. The total mass of the sphere is
\begin{equation}
m_1=\frac{4\pi}{3}\left[\rho_A R_A^3+\rho_B(R_B^3-R_A^3)+\rho_2(R_1^3-R_B^3)\right],
\end{equation}
so that the average density is
\begin{equation}
\label{A2}
\rho_1=\frac{m_1}{\frac{4\pi}{3}R_1^3}=\rho_A z_A^3+\rho_B(z_B^3-z_A^3)+\rho_2(1-z_B^3),
\end{equation}
where $z_A\equiv R_A/R_1$ and $z_B\equiv R_B/R_1$. Note that $0<z_A<z_B<1$. Equation \eqref{A2} expresses $\rho_1$ as a weighted average of $\rho_A$, $\rho_B$, and $\rho_2$, Obviously, $\min\{\rho_A,\rho_B,\rho_2\}\leq \rho_1\leq\max\{\rho_A,\rho_B,\rho_2\}$.

The moment of inertia of a spherical shell of density $\rho_B$ and radii $R_A$ and $R_B$ is $\frac{8\pi}{15}\rho_B (R_B^5-R_A^5)$. Thus, the moment of inertia of our sphere is
\begin{equation}
I_1=\frac{8\pi}{15}\left[\rho_A R_A^5+\rho_B(R_B^5-R_A^5)+\rho_2(R_1^5-R_B^5)\right],
\end{equation}
its reduced value being
\begin{equation}
\label{A4}
\kappa=\frac{I_1}{m_1R_1^2}=\frac{2}{5}\left[\frac{\rho_A}{\rho_1} z_A^5+\frac{\rho_B}{\rho_1}(z_B^5-z_A^5)+\frac{\rho_2}{\rho_1}(1-z_B^5)\right].
\end{equation}
Therefore, given $\rho_A/\rho_1$, $\rho_B/\rho_1$, $\rho_2/\rho_1$, and $\kappa$, Eqs.\ \eqref{A2} and \eqref{A4} allow one to obtain $z_A$ and $z_B$.

Henceforth, we assume that the  particle mimics a sphere with a uniform mass distribution, i.e., $\kappa=\frac{2}{5}$.
In that case,
Eqs.\ \eqref{A2} and \eqref{A4} can be rewritten as
\begin{equation}
\label{A6}
1=Y_{B2}z_B^3-Y_{BA}z_A^3=Y_{B2}z_B^5-Y_{BA}z_A^5,
\end{equation}
where
\begin{equation}
\label{A8}
Y_{B2}\equiv \frac{\rho_B-\rho_2}{\rho_1-\rho_2},\quad Y_{BA}\equiv \frac{\rho_B-\rho_A}{\rho_1-\rho_2} .
\end{equation}
It can be checked that the condition $0<z_A<z_B<1$ implies $1<Y_{B2}<1+Y_{BA}$, which yields
$\rho_A<\rho_1<\rho_B$ and $\rho_B<\rho_1<\rho_A$ for $\rho_1>\rho_2$ and  $\rho_1<\rho_2$, respectively.
For  simplicity, let us choose the same density for the inner core and the outer shell, i.e., $\rho_A=\rho_2$. In that case,
$Y_{BA}=Y_{B2}=Y$ and Eq.\ \eqref{A6} yields
\begin{equation}
\label{A11}
\frac{1}{Y}=z_B^3-z_A^3=z_B^5-z_A^5.
\end{equation}
In particular, if one chooses $Y=2$ the solution is $z_A=0.605\,907$, $z_B = 0.897\,293$.

In the case of a (big) intruder with $\sigma_1/\sigma_2=2$ and $m_1/m_2=2.721$ [see Fig.\ \ref{fig5}(a)], then $\rho_1/\rho_2=m_1\sigma_2^3/m_2\sigma_1^3=0.34$ and $(1-\rho_1/\rho_2)^{-1}=1.52$, so it is possible to choose  $Y=2$. In such a case, the density of the middle shell is $\rho_B/\rho_2=1-2\rho_1/\rho_2=0.32$.

Alternatively, for a (small) intruder with $\sigma_1/\sigma_2=\frac{1}{2}$ and $m_1/m_2=0.440$ [see Fig.\ \ref{fig5}(b)], $\rho_1/\rho_2=m_1\sigma_2^3/m_2\sigma_1^3=3.52$. If we again choose $Y=2$,  the density of the middle shell is $\rho_B/\rho_2=2\rho_1/\rho_2-1=6.04$.

\bibliography{D:/Dropbox/Mis_Dropcumentos/bib_files/Granular}

\end{document}